\documentclass{llncs}
\usepackage{amsmath,amsfonts,amssymb}
\usepackage{xspace}
\usepackage{url}
\usepackage{algorithm}
\usepackage[noend]{algpseudocode}
\usepackage{hyperref}
\usepackage{pgfplots}
\usepackage{xcolor}

\newcommand{\rmmr}{\emph{multi-round MapReduce scheduling} problem\xspace}
\newcommand{\smmr}{\emph{single-round MapReduce scheduling} problem\xspace}
\let\doendproof\endproof
\renewcommand\endproof{~\hfill\qed\doendproof}

\begin{document}

\title{Scheduling MapReduce Jobs under Multi-Round Precedences}

\author{
D. Fotakis\inst{1}
\and I. Milis\inst{2}
\and O. Papadigenopoulos\inst{1}
\and V. Vassalos\inst{2}
\and G. Zois\inst{2}
}

\institute{School of Electrical and Computer Eng., National Technical University of Athens\\
\email{fotakis@cs.ntua.gr}, \email{opapadig@corelab.ntua.gr}
\and Department of Informatics, Athens University of Economics and Business, Greece\\
\email{\{milis, vassalos, georzois\}@aueb.gr}}
\maketitle

\begin{abstract}
We consider non-preemptive scheduling of MapReduce jobs with multiple tasks in the practical scenario where each job requires several map-reduce rounds. We seek to minimize the average weighted completion time and consider scheduling on identical and unrelated parallel processors. For identical processors, we present LP-based $O(1)$-approximation algorithms. For unrelated processors, the approximation ratio naturally depends on the maximum number of rounds of any job. Since the number of rounds per job in typical MapReduce algorithms is a small constant, our scheduling algorithms achieve a small approximation ratio in practice. For the single-round case, we substantially improve on previously best known approximation guarantees for both identical and unrelated processors. Moreover, we conduct an experimental analysis and compare the performance of our algorithms against a fast heuristic and a lower bound on the optimal solution, thus demonstrating their promising practical performance.
\end{abstract}

\section{Introduction}

The sharp rise in Internet's 
use has boosted the amount of data stored on the web and processed daily.
MapReduce \cite{DEAN} and its open-source implementation Hadoop 
have been established as the standard platform for processing data sets on large clusters. 
The main idea behind MapReduce is inspired by the BSP model of parallel computation. 
A MapReduce job starts by allocating (randomly or arbitrarily) data to a set of processors. The computation over the dataset is broken into successive rounds, where, during each round, every map task of a job operates on a portion of the input data, translating it into a number of key-value pairs and, after an intermediate process, all pairs having the same key are transmitted to the same reduce task. Subsequently, the reduce tasks of the job operate on the values associated with the corresponding keys and generate a partial result. 
A key observation is that, while the map and reduce phases in each round must be executed sequentially, the tasks in each phase can be executed in parallel.

An algorithm's efficiency in the MapReduce model mostly depends on the local computation cost of map and reduce phases, on the communication cost of transmitting the intermediate data between map and reduce tasks and on the number of rounds.
Two models have been proposed for analyzing the efficiency of MapReduce algorithms.
Karloff et al.~\cite{KarloffSV10} presented a model inspired by PRAM and proved that a large class of PRAM algorithms can be efficiently implemented in MapReduce. For a MapReduce algorithm to be efficient, Karloff et al. require that the number of processors and their memory should be sublinear and the running time in each round should be polynomial in the input size. Given these restrictions, efficiency is mostly quantified by the number of rounds. Ideally, the number of rounds should be constant, which is indeed possible for many important algorithmic problems. 
Afrati et al.~\cite{ASSU13} proposed a different model that is inspired by BSP and focuses on the trade-off between communication and computation cost.
The main idea is that restricting the computation cost leads to a greater amount of parallelism and to a larger communication cost between the mappers and the reducers. 
In \cite{ASSU13}, the goal is to design algorithms that run in few rounds and minimize the communication cost, given an upper bound on the computation cost in the reducers.

Although MapReduce is a distributed computation model, the scheduler of a MapReduce system exploits the inherent parallelism by operating in a centralized manner. The design of high quality schedulers that optimize the total execution time with respect to various needs within a MapReduce computation has emerged as a challenging research topic. Such schedulers are crucial for the efficiency of large MapReduce clusters shared by many users. These clusters typically deal with many jobs that consist of many tasks and of several map-reduce rounds. In such processing environments, the quality of a schedule is typically measured by the jobs' average completion time, which for a MapReduce job takes into account the time when the last reduce task finishes its work. A few results have been recently presented in this context~\cite{MoseleyDKS11,ChenKL12,FM15},
in order to capture the main practical insights in a MapReduce computation (e.g., task dependencies, data locality), mainly based on variants of shop scheduling problems.

In this work, we present a general model and an algorithmic framework for scheduling non-preemptively a set of MapReduce jobs on parallel (identical or unrelated) processors under precedence constraints with the goal to minimize the average weighted completion time. We consider jobs consisting of multiple map-reduce rounds, where each round consists of multiple map and reduce tasks. Each reduce task cannot begin its execution before all map tasks of the same round are finished, while the same also holds between reduce and map tasks of two successive rounds. The tasks are associated with positive processing times, depending on the processor environment, and each job has a positive weight to represent its priority value.
Assuming positive values for the tasks' execution times, which are polynomially bounded by the input size, we are consistent with both computation models in \cite{KarloffSV10,ASSU13}. We refer to our problem as the \rmmr or the \smmr (depending on the number of rounds).

\smallskip\noindent{\bf Related Work.} As MapReduce is becoming the standard platform for more and more applications, new algorithmic questions arise. These deal with the distributed setting of MapReduce's architecture, where computations proceed in more than one map-reduce rounds and the main goal is to minimize the number of rounds (which is usually constant). Recent results in this direction \cite{MoseleySPAA13,ImM15} have proposed substantial improvements on the number of rounds for various MapReduce algorithms. In a slightly different direction, \cite{AJRSU15} presents multi-round MapReduce algorithms, trying to optimize the tradeoff between the communication cost and the number of rounds. 

Another line of research deals with the design of high quality schedulers. A significant volume of work concerns the experimental evaluation of scheduling heuristics, trying to achieve good tradeoffs between various criteria (see e.g., \cite{YooS11}). Theoretical work (e.g., \cite{MoseleyDKS11,ChenKL12,FM15}) focuses on scheduling a set of MapReduce jobs on parallel processors to minimize the average (weighted) completion time, in the restricted case where each job is executed in a single round. \cite{ChenKL12} presents approximation algorithms using simple models, equivalent to known variants of the open-shop problem, taking into account task precedences and assuming that the tasks are preassigned to processors. Moseley et al.~\cite{MoseleyDKS11} present a 12-approximation algorithm for the case of identical processors, modeling in this way MapReduce scheduling as a generalization of the so-called two-stage Flexible Flow-Shop problem. 
They also present a $O(1/\epsilon^2)$-competitive online algorithm, for any $\epsilon\in (0, 1)$, under $(1 + \epsilon)$-speed augmentation. \cite{FM15} studies the \smmr in the most general case of unrelated processors and present an LP-based 54-approximation algorithm. They also show how to incorporate the communication cost into their algorithm, with the same approximation ratio.

\smallskip\noindent{\bf Contribution.}
Our model incorporates all the main features of the 
models in \cite{KarloffSV10,ASSU13}, aiming at an efficient scheduling and assignment of tasks in MapReduce environments. Our contribution is threefold. First, in terms of modeling the MapReduce scheduling process: (i) We consider the practical scenario of multi-round multi-task MapReduce jobs and capture their task dependencies, and (ii) we study both identical and unrelated processors, thus dealing with data locality.
Second, in terms of algorithm design and analysis: (i) We propose an algorithmic framework for the \rmmr with proven performance guarantees, distinguishing between the case of indistinguishable and disjoint (map and reduce) sets of identical or unrelated processors, and (ii) our algorithms are based on natural LP relaxations of the problem and improve on the approximation ratios achieved in previous work~\cite{MoseleyDKS11,FM15}.
Third, in terms of experimental analysis, we focus on the most general case of unrelated processors and show that our algorithms have an excellent performance in practice. 

The rest of the paper is organized as follows. In Section 2, we formally define our model and provide notation.
In Section 3, we consider \rmmr on identical indistinguishable or disjoint processors. More precisely, in the
case of identical indistinguishable processors we propose a 4-approximation algorithm which is based on a reduction of the \rmmr to the classic scheduling of a set jobs on identical processors, under precedence constraints, so as to minimize their average weighted completion time~\cite{QueyranneS06}. Since the latter does not apply in the disjoint processor case, we further propose a novel 11-approximation algorithm that transforms a fractional schedule of an interval-indexed LP relaxation of the problem to an integral schedule by carefully
applying, on each interval of execution, a novel variant of the well-known Graham's list scheduling 2-approximation algorithm~\cite{Graham69} for scheduling a set jobs on identical processors, under precedence constraints, to minimize makespan. Moreover, for the \smmr on identical disjoint processors we substantially improve on the results proposed by Moseley et al.~\cite{MoseleyDKS11}, presenting an LP-based 8-approximation algorithm, instead of 12-approximation, by refining the general idea in~\cite{MoseleyDKS11,FM15} of creating partial schedules of only map and only reduce tasks and then merging these schedules losing a constant approximation factor. 

In Section 4, we consider the \rmmr on the most general environment of unrelated processors and
we propose a LP-based $O(r)$-approximation algorithm, where $r$ is the maximum number of rounds over all jobs. Our technique improves on ideas proposed in~\cite{FM15}, by formulating an interval-indexed LP relaxation of the \rmmr so as to handle the multi-round precedences.
Unlike previous work~\cite{MoseleyDKS11,FM15} we avoid the two-steps idea
of creating partial schedules of only map and only reduce tasks and then combine
them into one. As a result of this refinement, in the single-round case we show a $37.87$-approximation for the \smmr, substantially improving on the
previously proposed 54-approximation algorithm in~\cite{FM15}. Furthermore, we comment on the hardness of the \rmmr.

In Section 5, we compare our algorithms via simulations of random instances with a fast heuristic, proposed in~\cite{FM15}, as well as with a lower bound on the optimal value of the \rmmr. To capture data locality issues, we consider instances that use processor and task relations. The experiments show that our algorithm achieves a significantly better empirical ratio than the corresponding theoretical bound.

\section{Problem formulation}

\begin{figure}[t!]
  \centering
  \includegraphics[scale=0.55]{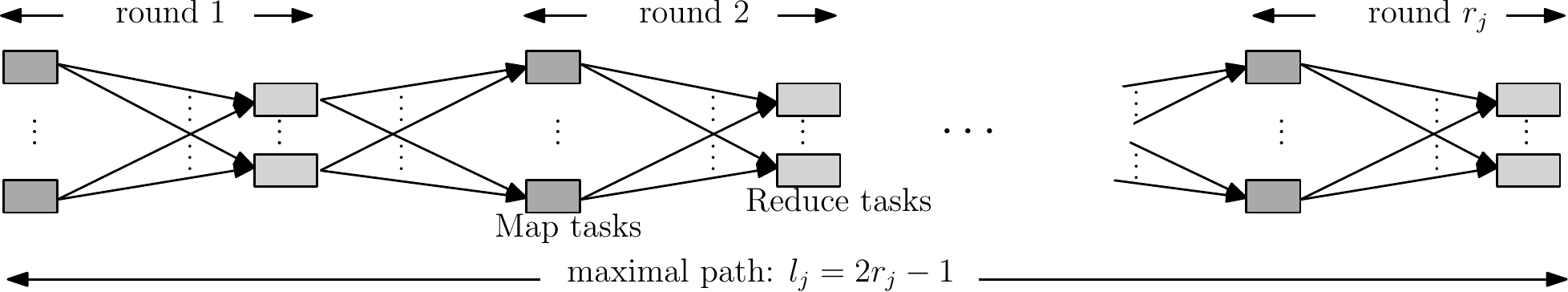}
  \caption{A MapReduce job $j$ of $r_j$ rounds, and length $l_j = 2r_j-1$.}\label{fig:1}
\end{figure}

We consider a set $\mathcal{J}=\{1,2,\ldots,n\}$ of $n$ \emph{MapReduce jobs} to be scheduled on a set $\mathcal{P}=\{1,2,\ldots,m\}$ of $m$ parallel \emph{processors}. Each job $j \in \mathcal{J}$ is available at time zero, has some positive weight $w_j$, and comprises of $r_j\in \mathbb{N},~r_j\geq 1$ rounds of computation, with each round consisting of a set of map tasks and a set of reduce tasks. Let $\mathcal{M}$, $\mathcal{R}$ be the sets of all \emph{map} and \emph{reduce} tasks respectively. 
Each task $T_{k,j}\in \mathcal{M}\cup\mathcal{R}$ of a job $j\in \mathcal{J}$, where $k \in N$, is associated with a positive processing time. Note that, by assuming task processing times that are polynomially bounded by the input size we are consistent with the two above computation models~\cite{ASSU13,KarloffSV10}. In every round, each reduce task can start its execution only after the completion of all map tasks of the same job, while the same precedence constraints hold also between the reduce and the map tasks of two successive rounds. In other words, except for the precedence constraints emerged by the existence of map and reduce phases, there are also precedence constraints between consecutive rounds, so in order for a map task of a round $r \in \{2, \dots r_j\}$, of a job $j$, to begin processing, all the reduce tasks of round $r-1$ have to complete their execution. A \emph{multi-round MapReduce job} $j$ can be represented by an {\it $r_j$-partite-like} directed acyclic graph, as the one depicted in Fig.~\ref{fig:1}, where $r_j$ is the number of rounds and $l_j = 2r_j -1$ is the length of a maximal path of the tasks' precedences. Throughout the analysis, in order to upper bound the approximation ratio of our algorithm, the latter parameter is used instead of the number of rounds.

To better capture data locality issues in task assignment, we distinguish between the standard \emph{identical processors} environment, where the processing time of each task $T_{k,j}$, let $p_{k,j}$, is the same on every processor, and the most general \emph{unrelated processors} environment, where there is a vector of processing times $\{p_{i,k,j}\}$, one for each processor $i\in \mathcal{P}$. Concerning the dedication of processors to either map or reduce tasks, we examine two cases: a) The sets $\mathcal{P_M}$ and $\mathcal{P_R}$ are \emph{indistinguishable} and the processors in $\mathcal{P}$ are processing both map and reduce tasks, and b) the set $\mathcal{P}$ is divided into two \emph{disjoint} sets $\mathcal{P_M}$ and $\mathcal{P_R}$, where $\mathcal{P} = \mathcal{P_M} \cup \mathcal{P_R}$, where the processors of $\mathcal{P_M}$ process only map tasks, while the processors of $\mathcal{P_R}$ process only reduce tasks.

For a given schedule we denote by $C_j$ and $C_{k,j}$ the completion times of a job $j \in \mathcal{J}$ and a task $T_{k,j}\in \mathcal{M}\,\cup\,\mathcal{R}$ respectively. Note that, due to the task precedences along the $r_j$ rounds of each job $j$, $C_j = \max_{T_{k,j}\in\mathcal{R}} \{C_{k,j}\}$. By $C_{max}=\max_{j \in \mathcal{J}} \{C_j\}$ we denote the \emph{makespan} of the schedule, i.e. the completion time of the last job. Our goal is to schedule \emph{non-preemptively} all tasks on processors of $\mathcal{P}$, with respect to their precedence constraints, so as to minimize the average weighted completion time, $\sum_{j\in\mathcal{J}} w_jC_j$.


%

%
%

%

\section{Scheduling Tasks on Identical Processors}

We next study \emph{multi-round MapReduce scheduling} on identical indistinguishable or disjoint processors. Reducing the problem to standard job scheduling under precedence constraints, we first obtain a 4-approximation algorithm for indistinguishable processors. This result holds also for \emph{single-round MapReduce scheduling}. Then, we present an 11-approximation algorithm for identical disjoint processors. In the same case, we propose an improved 8-approximation algorithm for the \smmr, which substantially improves on the 12-approximation algorithm proposed in \cite{MoseleyDKS11} for the same problem.  
\subsection{The case of indistinguishable processors}

We consider the \rmmr on identical indistinguishable processors.
Finding an algorithm for this problem can be easily reduced to finding an algorithm for the classic problem of scheduling a set of jobs on identical processors, under precedence constraints, to minimize their average weighted completion time. More specifically, for any instance of our problem we can create an equivalent instance of the latter problem: For every task $T_{k,j} \in \mathcal{M} \cup \mathcal{R}$, we create a corresponding job $j_k$ of equal processing time, $p_{k,j}$, and zero weight, $w_{j_k} = 0$. We maintain the same precedence constraints, emerged from the input of \rmmr, to the new problem, i.e. for every $T_{k,j} \succ T_{k',j}$ we set $j_k \succ j_{k'}$. For each MapReduce job $j$, we create a \emph{dummy} job $j_D$ of zero processing time and weight equal to the weight of $j$, i.e. $w_{j_D} = w_j$, and for every job $j_k$ we demand that $j_k \succ j_D$. In other words, since the corresponding dummy task of a MapReduce job $j$ has zero processing time it can be scheduled, in an optimal schedule, exactly after the completion time of all corresponding jobs $j_k$ and, therefore, indicate the completion time of the job itself in the MapReduce context. Moreover, every dummy job $j_D$ carries the weight of the corresponding MapReduce job $j$. 
\cite{QueyranneS06} shows a 4-approximation algorithm for scheduling a set of jobs on identical processors, under precedence constraints, to minimize their average weighted completion time. Combining our transformation with this algorithm, we obtain that:

\begin{theorem}
There is a 4-approximation algorithm for the \rmmr on identical indistinguishable processors.
\label{thm:ident-ind}
\end{theorem}

\subsection{The case of disjoint processors}

Inspired by the algorithm of \cite[Theorem~3.8]{HallSSW97}, we present an $O(1)$-approximation algorithm which transforms a solution to an interval-indexed LP relaxation of our problem into an integral schedule by carefully applying, on each interval of execution, a variation of the Graham's 2-approximation algorithm~\cite{Graham69} for job scheduling on identical processors under precedence constraints to minimize makespan.

We use the term $b \in \{\mathcal{M},\mathcal{R}\}$ to refer to both map and reduce attributes. Moreover, for any set of tasks $S \subseteq b$, we define $p(S) = \sum_{T_{k,j} \in S} p_{k,j}$ and $p^2(S) = \sum_{T_{k,j} \in S} p^2_{k,j}$.
The following (LP1) is an interval-indexed linear programming relaxation of our problem. Constraints (\ref{lpp:p1-1}) ensure that the completion time of a MapReduce job is at least the completion time of any of its tasks and that the completion time of any task is at least its processing time. Constraints (\ref{lpp:p2-2}) capture the relation of completion times of two tasks $T_{k,j} \succ T_{k',j}$. Constraints (\ref{lpp:p3-3}) have been proved \cite{HallSSW97} to hold for any feasible schedule on identical processors minimizing the average weighted completion time and give useful lower bounds to the completion times of tasks.

\begin{alignat}{2}
\text{{\bf (LP1)}:} \ \ \ \ \ \ &  \text{minimize} \sum_{j \in \mathcal{J}} w_j C_{j} \notag\\
\text{s.t.}\ \ \ \ \ \
& C_{j} \ge C_{k, j} \ge p_{k,j}\,~~~&\forall T_{k,j}\in \mathcal{M} \cup \mathcal{R}&\label{lpp:p1-1}\\
& C_{k, j} \ge C_{k', j} + p_{k, j}\,~~&\forall T_{k',j} \prec T_{k,j} &\label{lpp:p2-2}\\
& \sum_{T_{k,j} \in b} p_{k,j} C_{k,j} \ge {{p(S)^2 + p^2(S)} \over 2|\mathcal{P}_b|}\,\ \ \  \
~~&b \in \{\mathcal{M}, \mathcal{R}\}, \forall S \subseteq b &\label{lpp:p3-3}
\end{alignat}

Let $(0,t_{\max} = \sum_{T_{k,j}\in \mathcal{M} \cup \mathcal{R}} p_{k,j}]$ be the time horizon of the schedule, where $t_{\max}$ is an upper bound on the makespan of any feasible schedule. We discretize the time horizon into intervals $[1,1],(1,2], (2,2^2],\ldots,(2^{L-1},2^L]$, where $L$ is the smallest integer such that $2^{L-1} \geq t_{\max}$. Let $\mathcal{L} = \{1,2,\ldots, L\}$. Note that, interval $[1,1]$ implies that no job finishes its execution before time 1; in fact, we can assume, w.l.o.g., that all processing times are positive integers. Let $\tau_0 = 1$ and $\tau_\ell = 2^{\ell - 1}$. Our algorithm begins from a fractional solution to the LP, $(\bar{C}_{k,j}, \bar{C}_j)$, and separates tasks into intervals with respect to their completion times $\bar{C}_{k,j}$ as follows.

Let $S(\ell) = \{ T_{k,j} | \tau_{\ell-1} < \bar{C}_{k,j} \leq \tau_{\ell}  \}$. Let also $S^M(\ell) \subseteq S(\ell)$ and $S^R(\ell) \subseteq S(\ell)$ be a partition of each set $S(\ell)$ into only map and only reduce tasks, respectively. We define $t^M_{\ell}= {p(S^M(\ell)) \over |\mathcal{P}_M|}$ and $t^R_{\ell} = {p(S^R(\ell)) \over |\mathcal{P}_R|}$ to be the average load of a map and reduce processor, respectively, of the map and reduce tasks of each set $S(\ell)$. Now, we can define an adjusted set of intervals as $\bar{\tau}_{\ell} = 1 + \sum^{\ell}_{k=1}(\tau_k + t^M_{k}  + t^R_{k})  \quad \forall \ell \in \mathcal{L}$.
We can schedule greedily the tasks of each set $S(\ell)$ in interval $(\bar{\tau}_{\ell-1}, \bar{\tau}_{\ell}]$, using the following variation of Graham's List Scheduling algorithm.

\begin{quote}
\textsc{Two-Resource List Scheduling.} Consider two different types of available resources, i.e. the map and the reduce processors, while each task can be scheduled only on a specific resource type. Whenever a processor becomes available, execute on it any available unscheduled task that corresponds to its type.
\end{quote}

\begin{lemma}
The tasks of $S(\ell)$ can be scheduled non-preemptively at interval $(\bar{\tau}_{\ell-1}, \bar{\tau}_{\ell}]$ by applying \textsc{Two-Resource List Scheduling}.
\end{lemma}
\begin{proof}
Using the analysis of \cite{HallSSW97} we can prove that the makespan of each set $S(\ell)$ is upper bounded by the sum of the longest chain of precedences, the average processing time of a map processor and the average processing time of a reduce processor. By definition of $S(\ell)$ and constraints (\ref{lpp:p1-1}), we know that the the longest chain of precedences can be at most $\tau_{\ell}$. Therefore, if the algorithm starts by assigning tasks at time $\bar{\tau}_{\ell-1}$, it should have finished by time $C \leq \bar{\tau}_{\ell-1} + \tau_{\ell} + t^M_{\ell}+t^R_{\ell}$. Then, by definition of $\bar{\tau}_{\ell-1}$ we have that $C \leq 1 + \sum^{\ell - 1}_{k=1}(\tau_k + t^M_{k}  + t^R_{k}) + \tau_{\ell} + t^M_{\ell}+t^R_{\ell} \leq 1 + \sum^{\ell}_{k=1}(\tau_k + t^M_{k}  + t^R_{k}) = \bar{\tau}_{\ell}$.\\[1ex]
\end{proof}

Note that the resulting schedule respects the tasks precedences since by (\ref{lpp:p1-1}), for any pair of tasks such that $T_{k,j} \succ T_{k',j}$, it must be the case that $T_{k,j} \in S(\ell)$ and $T_{k',j} \in S(\ell')$ with $\ell \leq \ell'$. Now we are able to prove the following theorem.

\begin{theorem}
There is an 11-approximation algorithm for the \rmmr on identical disjoint processors.
\end{theorem}
\begin{proof}
Consider the completion time $C_{k,j}$ of a task $T_{k,j} \in S(\ell)$. By constraints (\ref{lpp:p1-1}) and (\ref{lpp:p2-2}) we know that the length of any chain that ends with $T_{k,j}$ is upper bounded by $\bar{C}_{k,j}$. Therefore, using the previous lemma and since $T_{k,j} \in S(\ell)$, we can see that for its completion time it holds:
\begin{align*}
C_{k,j} \leq \bar{\tau}_{\ell-1} + t^M_{\ell} + t^R_{\ell} + \bar{C}_{k,j} \\
= 1 + \sum^{\ell -1 }_{k=1} (\tau_k + t^M_{k} + t^R_{k}) + t^M_{\ell} + t^R_{\ell} + \bar{C}_{k,j}\\
= \tau_\ell + \sum^{\ell}_{k=1}( t^M_{k} + t^R_{k}) + \bar{C}_{k,j}
\end{align*}
Constraints (\ref{lpp:p3-3}) imply that, let $T_{k',j'}$ the last finishing -say map- task of the set $S(\ell')$, it holds $\bar{C}_{k',j'} \geq {1 \over 2 |\mathcal{P}_M|} \sum^{\ell'}_{k=1} p(S(k))$. The same holds for the reduce tasks.
Therefore:
\begin{align*}
\sum^{\ell}_{k=1}( t^M_{k} + t^R_{k}) = \sum^{\ell}_{k=1} t^M_{k} + \sum^{\ell}_{k=1}t^R_{k}\\
\leq {1 \over |\mathcal{P}_M|} \sum^{\ell}_{k=1}p(S^M(k)) + {1 \over |\mathcal{P}_R|} \sum^{\ell}_{k=1}p(S^R(k)) \\
\leq 2 \tau_\ell + 2 \tau_\ell = 4 \tau_\ell
\end{align*}
Since by definition of $S(\ell)$, $\tau_{\ell} \leq 2 \bar{C}_{k,j}$ it is the case that:
\begin{align*}
C_{k,j} \leq  \tau_\ell + 4 \tau_\ell + \bar{C}_{k,j} \leq 11 \bar{C}_{k,j}
\end{align*}
The theorem follows by applying the previous inequality to the objective function.\\[1ex]
\end{proof}
\noindent\emph{Remark.} A simple transformation of the previous algorithm yields a 7-approximation algorithm for indistinguishable processors. However, Theorem \ref{thm:ident-ind} also applies and gives a 4-approximation algorithm for the \smmr.

\subsection{The single-round case}

For \emph{single-round MapReduce scheduling}, we obtain an 8-approximation algorithm. Our algorithm improves on the 12-approximation algorithm of \cite{MoseleyDKS11,FM15},  while refines their idea (of merging independent schedules of only map and only reduce tasks on their corresponding sets of processors into a single schedule) by applying a 2-approximation algorithm similar to that in~\cite[Lemma~6.1]{CorreaSV12}. Note that \cite{CorreaSV12} considers a set of orders of jobs, instead of jobs consisting of tasks, and the completion time of each order is specified by the completion of the job that finishes last.

\begin{alignat}{2}
\text{{\bf (LP2)}:} \ \ \ \ \ \  &\text{minimize} \sum_{j \in \mathcal{J}} w_j C_{j} \notag\\
\text{s.t.}  \ \ \ \ \ \ & C_j \geq M_{k,j} + {p_{k,j}\over 2}\,~~~~~~~~~&\forall T_{k,j} \in b&\label{lps:p1-1}\\
& \sum_{T_{k,j} \in S} p_{k,j} M_{k,j} \geq {p(S)^2 \over 2|\mathcal{P}_b|}\,~~~~~~&\forall S \subseteq b &\label{lps:p2-2}
\end{alignat}

For the partial schedules $\sigma_b$ 
of only map and only reduce tasks, since we have no precedence constraints between tasks, let $M_{k,j}$ be the midpoint of a task $T_{k,j} \in b$ in any non-preemptive schedule, i.e., $M_{k,j} = C_{k,j} - {p_{k,j}\over 2}$. \cite{EastmanEI64} shows that in any feasible schedule on $m$ identical processors, for every $S \subseteq b :$ $\sum_{T_{k,j} \in S} p_{k,j} M_{k,j} \geq {p(S)^2 \over 2m}$.

Now, consider the linear programming formulation (LP2). Note that, although the number of inequalities of this linear program is exponential, it is known \cite{QUE} that it can be solved in polynomial time using the ellipsoid algorithm. Thus, consider an optimal solution $(\bar{M}_{k,j}, \bar{C}_j)$ to this formulation with objective value $\sum_{j \in \mathcal{J}} w_j \bar{C}_j$. If we greedily assign tasks on the processors of $\mathcal{P}_b$ in a non-decreasing order of $\bar{M}_{k,j}$ using Graham's list scheduling, then, for the resulting schedule $\sigma_b$, it holds that:

\begin{lemma}
There is a 2-approximate schedule of map (resp. reduce) tasks on identical map (resp. reduce) processors to minimize their average weighted completion time.
\label{le:2-appr.}
\end{lemma}
\begin{proof}
Consider a task $T_{k,j}$ and let $C_{k,j}$ be its completion time in the produced schedule. We denote by $\mathcal{B}$ the set of tasks that are scheduled before $T_{k,j}$ by the list scheduling algorithm. By definition, the midpoint value of any task in $\mathcal{B}$ is at most $\bar{M}_{k,j}$. Let $S_{k,j}$ be the starting time of $T_{k,j}$ in the schedule. Given that by time $S_{k,j}$ all processors are busy, it certainly holds that $S_{k,j} \leq {p(\mathcal{B}) \over |\mathcal{P}_b|}$. Therefore, using the constraints of the LP formulation and since $\bar{M}_{k,j}$ is greater of equal to the midpoints of every task in $\mathcal{B}$, we can see that:
\begin{align*}
\bar{M}_{k,j} p(\mathcal{B} \cup \{T_{k,j}\}) \geq \sum_{T_{k',j'} \in \mathcal{B} \cup \{T_{k,j}\}} p_{k',j'} \bar{M}_{k',j'} \geq {p(\mathcal{B} \cup \{T_{k,j}\})^2 \over 2|\mathcal{P}_b|}
\end{align*}
Therefore, since $\bar{M}_{k,j} \geq {p(\mathcal{B} \cup \{T_{k,j}\}) \over 2|\mathcal{P}_b|}$, we have that:
\begin{align*}
C_{k,j} &= S_{k,j} + p_{k,j} \leq {p(\mathcal{B}) \over |\mathcal{P}_b|} + p_{k,j} \leq {p(\mathcal{B} \cup \{T_{k,j}\}) \over |\mathcal{P}_b|} + p_{k,j} \\
&\leq 2 \bar{M}_{k,j} + p_{k,j} \leq 2(\bar{M}_{k,j} + {p_{k,j} \over 2}) \leq 2 \bar{C}_{k,j}
\end{align*}
\par Now, since $C_j = \max_{T_{k,j}} {C_{k,j}}$, it is the case that $C_j \leq 2 \bar{C}_j$.\\[1ex]
\end{proof}

The second step of our algorithm is to merge the two partial schedules $\sigma_M$ and $\sigma_R$ into a single one. To succeed it, we can use the merging technique proposed in \cite{MoseleyDKS11}. If we denote by $C^{\sigma_M}_j$ and $C^{\sigma_R}_j$ the completion times of a job $j$ in $\sigma_M$ and $\sigma_R$ respectively, we can define the \emph{width} of each job $j$ to be $\omega_j = \max\{C^{\sigma_M}_j, C^{\sigma_R}_j\}$. The algorithm schedules the tasks of each job on the same processors that they have been assigned in $\sigma_M$ and $\sigma_R$, in non-decreasing order of $\omega_j$.

\begin{theorem}
There is a 8-approximation algorithm for the \smmr on identical disjoint processors.
\end{theorem}
\begin{proof}
Let $OPT^M$ and $OPT^R$ be the optimal values of the partial problems of scheduling only map and only reduce tasks respectively. Let also $C^{\sigma_M}_j$ and $C^{\sigma_R}_j$ be the completion time of a job in each of these partial schedules. If we denote by $OPT$ the optimal value of our problem and by the Lemma~\ref{le:2-appr.} we have that:
\begin{align*}
\sum_{j \in \mathcal{J}} w_j C^{\sigma_b}_j \leq 2 OPT^b \leq 2 OPT
\end{align*}
Consider a task $j$ and let $C_j$ be its completion time in the final schedule. One can easily see \cite{MoseleyDKS11} that it is the case: $C_j \leq 2 \omega_j \leq 2(C^{\sigma_M}_j + C^{\sigma_R}_j)$.
Therefore, for the objective value of our schedule it holds:
\begin{align*}
\sum_{j \in \mathcal{J}} w_j C_j \leq \sum_{j \in \mathcal{J}} 2 w_j \omega_j \leq \sum_{j \in \mathcal{J}} 2 w_j (C^{\sigma_M}_j + C^{\sigma_R}_j)  \leq 8 OPT
\end{align*}\\[1ex]
\end{proof}

\noindent \emph{Remark.} The same analysis yields an 8-approximation algorithm for \emph{single-round MapReduce scheduling} on identical indistinguishable processors. We only have to define the width of each job to be $\omega_j = C^{\sigma_M}_j + C^{\sigma_R}_j$.

\section{Scheduling Tasks on Unrelated Processors}

In this section, we consider the \rmmr on unrelated processors. We present a $\mathcal{O}(l_{\max})$-approximation algorithm, where $l_{\max} = \max_{j \in \mathcal{J}} l_j$ is the maximum length over all jobs' maximal paths in the underlying precedence graph. Since $l_{max} = 2r_{\max} - 1$, our algorithm is also a $O(r_{\max})$-approximation, where $r_{\max}$ is the maximum number of rounds over all jobs. Our technique builds on ideas proposed in \cite{FM15}. We formulate an interval-indexed LP relaxation for \emph{multi-round MapReduce scheduling} so as to handle the multi-round precedences. Unlike \cite{MoseleyDKS11,FM15}, we avoid the idea of creating partial schedules of only map and only reduce tasks and then combine them into one. Applying the above algorithm for the \smmr, we derive a 37.87-approximation algorithm, thus improving on the 54-approximation algorithm of \cite{FM15}. Even though in the following analysis, we consider the case of indistinguishable processors, we can simulate the case of disjoint processors by simply setting $p_{i,k,j} = +\infty$ for every map (resp. reduce) task $T_{k,j}$ when $i$ is a reduce (resp. map) processor. In the sequel, we denote by $\mathcal{T} = \mathcal{M}\cup \mathcal{R}$ the set of all tasks.

\begin{alignat}{2}
\text{{\bf (LP3)}:} \ \ \ \ \ \  & \text{minimize} \sum_{j \in \mathcal{J}} w_j C_{j} \notag\\
\text{s.t.}  \ \ \ \ \ \ &
\sum_{i\in\mathcal{P}, \ell\in \mathcal{L}} y_{i, k, j, \ell} \geq 1,~~~~~~~~~~~~~~~~~~~~~~~~~&\forall T_{k,j}\in \mathcal{T}&\label{lp:p1-1}\\
 & C_{j} \ge C_{k, j},\,~~~~~~~~~~~~~~~~~~~~~~~~~~~~~~~&\forall T_{k,j}\in \mathcal{T}&\label{lp:p2-2}\\
 & C_{k, j} \ge C_{k', j} + \sum_{i\in \mathcal{P}} p_{i, k, j} \sum_{\ell\in \mathcal{L}} y_{i, k, j, \ell} ,\,~&\forall T_{k',j} \prec T_{k,j} &\label{lp:p3-3}\\
 & \sum_{i\in\mathcal{P}}\sum_{\ell\in \mathcal{L}} (1 + \delta)^{\ell-1} y_{i, k, j, \ell} \le C_{k, j},~~~~~~~~~~&\forall T_{k,j}\in \mathcal{T}\label{lp:p4-4}\\
 & \sum_{T_{k, j}\in \mathcal{T}} p_{i, k, j} \sum_{t \le \ell} y_{i, k, j, t} \le (1 + \delta)^\ell,~~~~~&\forall i \in \mathcal{P}, \ell\in \mathcal{L}\label{lp:p5-5}\\
 & p_{i, k, j} > (1 + \delta)^{\ell} \Rightarrow y_{i, k, j, \ell} = 0,~&\forall i \in \mathcal{P}, T_{k,j}\in b, \ell\in \mathcal{L}\label{lp:p6-6}\\
 & y_{i, k, j, \ell} \ge 0,~~~~~~~~~~~~~~~~~~~~~&\forall i \in \mathcal{P}, T_{k,j}\in b, \ell\in \mathcal{L}\notag
\end{alignat}

We use an interval-indexed LP relaxation.
Let $(0,t_{\max} = \sum_{T_{k,j}\in \mathcal{T}}\max_{i\in \mathcal{P}}p_{i,k,j}]$ be the time horizon of potential completion times, where $t_{\max}$ is an upper bound on the makespan of any feasible schedule.  Similarly with (LP1), we discretize the time horizon into intervals $[1,1],(1,(1+\delta)], ((1+\delta),(1+\delta)^2],\ldots,((1+\delta)^{L-1},(1+\delta)^L]$, where $\delta\in (0,1)$ is a small constant, and $L$ is the smallest integer such that $(1+\delta)^{L-1} \geq t_{\max}$.
Let $I_{\ell}=((1+\delta)^{{\ell}-1},(1+\delta)^{\ell}]$, for $1 \leq \ell\leq L$, and $\mathcal{L} = \{1,2,\ldots, L\}$.
Clearly, the number of intervals is polynomial in the size of the instance and in $1 \over \delta$.

We introduce an assignment variable $y_{i,k,j,\ell}$ indicating whether task $T_{k,j} \in \mathcal{T}$ is completed on processor $i \in \mathcal{P}$ within the interval $I_{\ell}$. Furthermore, let $C_{k,j}$ be the completion time variable for a task $T_{k,j} \in \mathcal{T}$ and $C_j$ be the completion time variable for a job $j\in \mathcal{J}$. Moreover, let $T_{k,j} \prec T_{k',j}$ be the precedence relation between two tasks. (LP3) is a LP relaxation of the \rmmr, whose corresponding integer program is itself a $(1+\delta)$-relaxation.

Constraints~(\ref{lp:p1-1}) ensure that every task is completed on a processor of the set $\mathcal{P}$ in some time interval. Constraints~(\ref{lp:p2-2}) denote that the completion time of a job is determined by the completion time of its last finishing task. Constraints~(\ref{lp:p3-3}) describe the relation between the completion times of two jobs $T_{k,j} \succ T_{k',j}$, where the term $\sum_{i\in \mathcal{P}} p_{i, k, j} \sum_{\ell\in \mathcal{L}} y_{i, k, j, \ell}$ refers to the fractional processing time of $T_{k,j}$. Constraints~(\ref{lp:p4-4}) impose a lower bound on the completion time of each task. For each $\ell\in \mathcal{L}$, constraints (\ref{lp:p5-5}) and (\ref{lp:p6-6}) are validity constraints which state that the total processing time of jobs executed up to an interval $I_{\ell}$ on a processor $i\in\mathcal{P}$ is at most $(1+\delta)^{\ell}$, and that if processing a task $T_{k,j}$ on a processor $i\in \mathcal{P}$ is greater than $(1+\delta)^{\ell}$, $T_{k,j}$ should not be scheduled on $i$, respectively.

\begin{algorithm}[t!]
\caption{\label{alg:mrs}\textsc{Multi-round MRS}: An algorithm for \emph{multi-round MapReduce scheduling} on unrelated processors}
\begin{algorithmic}[1]
\State Compute a fractional solution to the LP $(\bar{y}_{i,k,j,\ell}, \bar{C}_{k,j}, \bar{C}_{j})$.
\State Partition the tasks into sets $S(\ell) = \{T_{k,j}\in b~|~(1+\delta)^{\ell-1} \leq \alpha \bar{C}_{k,j} < (1+\delta)^{\ell} \}$, \\where $\alpha>1$ is a fixed constant.
\For {each $\ell = 1 \dots L$}
\If {$S(\ell) \neq \emptyset$}
\State Let $G_{\ell}$ be the precedence graph of the tasks of $S(\ell)$.
\State $V_{1, \ell}, \dots, V_{t, \ell}, \dots, V_{l_{\max}+1, \ell} \leftarrow$ \textsc{Decompose($G_{\ell}$)}
\For {each $V_{t, \ell}$, in increasing order of $t$}
\State Integrally assign the tasks of $V_{t,\ell}$ on $\mathcal{P}$ using~\cite[Theorem 2.1]{ST93}.
\State Schedule tasks of $V_{t, \ell}$ on $\mathcal{P}$, as early as possible, w.r.t. their precedences.
\EndFor
\EndIf
\EndFor
\end{algorithmic}
\end{algorithm}

Algorithm~\ref{alg:mrs} considers a fractional solution $(\bar{y}_{i,k,j,\ell}, \bar{C}_{k,j}, \bar{C}_{j})$ to (LP3) and rounds it to an integral schedule. It begins by separating the tasks into disjoint sets $S(\ell), \ell \in \mathcal{L}$ according to their fractional completion times $\bar{C}_{k,j}$. Since some of the tasks of each $S(\ell)$ may be related with precedence constraints, we proceed into a further partitioning of each set $S(\ell), \ell \in \mathcal{L}$ into pairwise disjoint sets $V_{t,\ell}, 1 \leq t \leq l_{\max}+1$, with the following property: all the predecessors of any task in $V_{t,\ell}$ must belong either in a set $V_{t',\ell}$ with $t' < t$, or in a set $S(\ell')$ with $\ell' <\ell$. Let $G$ be the precedence graph, given as input of the \rmmr. The above partitioning process on $G$ can be done in polynomial time by the following simple algorithm.

\begin{quote}
\textsc{Decompose($G$)}. Identify the nodes of zero out-degree, i.e., $\delta^-(v) = 0$, in $G$. Add them in a set $V_{t, \ell}$, starting with $t=1$, and remove them from the graph. Repeat until there are no more nodes. Output the sets of tasks.
\end{quote}

As the maximum path length in the precedence graph is $l_{\max}$, for each $\ell \in \mathcal{L}$, we could have at most $l_{\max}+1$ sets $V_{t, \ell}$, with some of them possibly empty. Now, since there are no precedence constraints among the tasks of each set $V_{t, \ell}$, we integrally assign these tasks using the algorithm of \cite[Theorem 2.1]{ST93} in an increasing order of $\ell$ and $t$.
The next lemmas prove an upper bound on the integral makespan of the tasks of every set $S(\ell)$ and $V_{t,\ell}$.

\begin{lemma} \label{le:lb1}
Suppose that we ignore any possible precedences among the tasks in $S(\ell)$, for each $\ell \in \mathcal{L}$. Then we can (fractionally) schedule them on the processors $\mathcal{P}$ with makespan at most ${\alpha \over \alpha-1} (1 + \delta)^\ell$.
\end{lemma}
\begin{proof}
We are going to apply the filtering technique proposed by Lin et al. \cite{LinV92}. First, we need to prove that, for any task $T_{k,j} \in S(\ell)$ it must hold that: $\sum_{i \in \mathcal{P}} \sum_{t \geq \ell +1} \bar{y}_{i,k,j,t} \leq {1\over \alpha}$. Suppose, through contradiction, that this is not the case and $\sum_{i \in \mathcal{P}} \sum_{t \geq \ell +1} \bar{y}_{i,k,j,t} > {1\over \alpha}$. Then, from (\ref{lp:p4-4}) we have $ \bar{C}_{k,j} \geq \sum_{i \in \mathcal{P}} \sum_{\ell \in \mathcal{L}} (1+\delta)^{\ell-1} \bar{y}_{i,k,j,\ell} > {1\over \alpha}(1+\delta)^{\ell}$, leading to a contradiction to the definition of $S(\ell)$. Given this, from (\ref{lp:p1-1}), it must be the case that $\sum_{i \in \mathcal{P}} \sum_{t \leq \ell} \bar{y}_{i,k,j,t} \geq {\alpha-1 \over \alpha}$. Now, we can transform the fractional solution $\bar{y}$ into a solution $y^*$ by setting for every task $T_{k,j} \in S(\ell)$ $y^*_{i,k,j,t} = 0$ for $t \geq \ell+1$ and $y^*_{i,k,j,t} = {\bar{y}_{i,k,j,t} \over \sum_{i \in \mathcal{P}} \sum_{t' \leq \ell} \bar{y}_{i,k,j,t'}}$ for $t \leq \ell$. We can easily verify that for the transformed solution $y^*$, constraints (\ref{lp:p1-1}), (\ref{lp:p4-4}) and (\ref{lp:p5-5}) are satisfied, if we multiply the right-hand side of the inequality with ${\alpha \over \alpha -1}$. The lemma follows from the fact that,  when we ignore the precedences among the tasks in $S(\ell)$, constraints (\ref{lp:p5-5}) indicate an upper bound to their fractional makespan.\\[1ex]
\end{proof}

Now, since every set of tasks $V_{t,\ell}$ is a subset of $S(\ell)$, the aforementioned result on the fractional makespan of $S(\ell)$ also holds for every $V_{t,\ell} \subseteq S(\ell)$.

\begin{lemma}
The tasks of every set $V_{t,\ell} \subseteq S(\ell)$ can be integrally scheduled on the processors $\mathcal{P}$ with makespan at most $({\alpha \over \alpha -1} + 1)(1 + \delta)^{\ell}$.
\label{le:lb2}
\end{lemma}
\begin{proof}
As already-mentioned, since the set $V_{t,\ell}$ is a result of the decomposition of the precedence graph of $S(\ell)$, the tasks that belong to it do not have precedence constraints among them. Moreover, the fractional load of these tasks on every processor follows the load of $S(\ell)$ and, by constraints (\ref{lp:p6-6}), the maximum processing time of any task in $V_{t,\ell}$ can be at most $(1+\delta)^{\ell}$. Therefore, using the rounding theorem~\cite[Theorem 2.1]{ST93}, we can integrally assign the tasks of $V_{t,\ell}$ on $\mathcal{P}$ with makespan at most $({\alpha \over \alpha -1} + 1)(1 + \delta)^{\ell}$.\\[1ex]
\end{proof}

Consider now a set of tasks $S(\ell)$ whose decomposition results in a sequence of pairwise disjoint subsets $V_{1, \ell}, \dots, V_{t, \ell}, \dots, V_{l_{\max}+1, \ell}$. Using the Lemma~\ref{le:lb2}, we see that if we integrally schedule each subset $V_{t, \ell}$ in a time window of $({\alpha \over \alpha -1} + 1)(1 + \delta)^{\ell}$ and then place the schedules in an increasing order of $t$, the resulting schedule would respect all constraints and would have makespan at most $(r+1)({\alpha \over \alpha -1} + 1)(1 + \delta)^{\ell}$. Now, we can prove the following.

\begin{theorem}
Algorithm~\ref{alg:mrs} is an $\alpha [(l_{\max}+1) {\alpha \over \alpha - 1} + l_{\max} {\alpha \over \delta (\alpha - 1)} + l_{\max} + 1 + {l_{\max}+ 1 \over \delta}] (1 + \delta)$-approximation for the \rmmr on unrelated processors, where $l_{\max}$ is the maximum length over all maximal paths in the precedence graph, and $\alpha > 1$, $\delta >0$ are fixed constants.
\label{thm:unrelated}
\end{theorem}
\begin{proof}
First, we need to note that the tasks of each set $S(\ell)$ can be scheduled integrally in the processors of $\mathcal{P}$ with makespan equal to the sum of makespans of the subsets $V_{t, \ell}, 1 \leq t \leq l_{\max}+1$. The rounding theorem of \cite[Theorem 2.1]{ST93} suggests that the makespan of an integral schedule of tasks in $V_{t,\ell}$ is at most the fractional assignment, $\Pi_{t, \ell} \leq {\alpha \over \alpha -1} (1+ \delta)^\ell$, of tasks to processors plus the maximum processing time on every processor, $p^{max}_{t, \ell} \leq (1+\delta)^\ell$. Therefore, the sets $V_{1,\ell}$ to $V_{l_{\max},\ell}$ can be scheduled with makespan at most $l_{\max} ({\alpha \over \alpha -1} + 1)(1 + \delta)^{\ell}$, in order to respect the precedences among them. Now, consider the sets $V_{l_{\max}+1, \ell}, \forall \ell \in \mathcal{L}$. Clearly, these must include the last finishing tasks of any chain in the precedence graph. Therefore, by constraints (\ref{lp:p5-5}), it is the case that $\sum_{t\leq \ell} \Pi_{l_{\max}+1,t} \leq {\alpha \over \alpha -1} (1+ \delta)^\ell$.
\par
Now, let $T_{k,j} \in \mathcal{T}$ be the last finishing task of a job $j \in \mathcal{J}$ which is scheduled on a processor $i\in \mathcal{P}$. Suppose, w.l.o.g., that $T_{k,j}$ belongs to the set $S(\ell)$. By Lemma ~\ref{le:lb2} and Lemma~\ref{le:lb1}, taking the union of the schedules of tasks in $S(\ell')$, with $\forall \ell' \leq \ell$, it must hold that the completion time of $T_{k,j}$ in the resulting schedule is:
\begin{align*}
C_{k,j} &\leq \sum_{\ell'\leq \ell} [l_{\max}({a \over a -1} + 1)(1+\delta)^{\ell'} + \Pi_{l_{\max}+1, {\ell'}} + p^{max}_{l_{\max}+1, {\ell'}}] \\
&\leq \alpha \left((l_{\max}+1) {\alpha \over \alpha - 1} + l_{\max} {\alpha \over \delta (\alpha - 1)} + l_{\max} + 1 + {l_{\max}+ 1 \over \delta} \right) (1 + \delta) \bar{C}_{k,j}.
\end{align*}
\end{proof}

As for \emph{single-round MapReduce scheduling}, for all the maximal paths of each job $j$ in the underlying graph, $l_j = 1$. By Theorem~\ref{thm:unrelated} with $(\alpha,\delta) \approx (1.65, 0.80)$, we get that:

\begin{corollary}
There is a $37.87$-approximation algorithm for the \smmr on unrelated processors.
\end{corollary}

\noindent{\bf A Note on the Computational Complexity.} Concerning the hardness of the \rmmr on unrelated processors, using an argument similar to the one of Section 3.1, we can easily verify that for the case of classic job scheduling on unrelated processors, under precedence constraints, a constant approximation algorithm for the problem of minimizing the average weighted completion time, implies a constant algorithm for the makespan objective.
Despite the importance and generality of the latter problem, to the best of our knowledge, the most general results that have been proposed so far concern the special cases where the underlying undirected precedence graph forms a set of chains \cite{ShmoysSW94} or forest (a.k.a. treelike precedence constraints)~\cite{KumarMPS09}, resulting in polylogarithmic approximation algorithms. 
In the \rmmr on unrelated processors the undirected graph underlying the precedence constraints forms a forest of $r$-partite-like graphs, where $r$ is the maximum number of rounds. Thus, it strictly generalizes on the chain precedences, since for the makespan objective, the fact that each job consists of a set of tasks does not affect the quality of a schedule.
We can also observe that the \rmmr is generalizing on the standard job-shop scheduling (where the precedence are restricted to be a disjoint union of chains and the task assignment is given in advance) under the same objective. However, for the latter one, we know that it is $\mathcal{NP}-hard$ to obtain a constant approximation algorithm and has no $O((\log lb)^{1-\epsilon})$-approximation algorithm for any $\epsilon > 0$, unless $NP\subseteq ZTIME({2^{\log n}}^{O(\frac{1}{\epsilon})})$, where $lb$ is a standard lower bound on the makespan of any schedule~\cite{MastrolilliS11}. Thus, the best we can expect is no more than a logarithmic improvement on our approximation ratio.

\section{Simulation results}

We conclude with simulation results for \emph{multi-round MapReduce scheduling} on unrelated processors. We compare our algorithm against the simple heuristic Fast-MR of \cite{FM15} and against a lower bound derived from (LP3). We provide evidence that the empirical approximation ratio of Algorithm~\ref{alg:mrs} is significantly better than the theoretical one.

Fast-MR operates in two steps. First, it computes an online assignment of tasks to processors, using the online algorithm of~\cite{AspnesAFPW97}, and then, it schedules them using a variant of Weighted Shortest Processing Time first wrt. the multi-round task precedences.

\smallskip\noindent{\bf Computational Experience and Results.} We generate instances consisting of 30 indistinguishable processors and from 5 to 50 jobs. Each job consists of 5 rounds, where the number of map and reduce tasks in each round ranges from $20$ to $35$ and from $5$ to $15$, respectively. The weight of each job is uniformly distributed in $[1, n]$, where $n$ is the number of jobs. Moreover, the parameters of Algorithm~\ref{alg:mrs} are fixed to $\delta = 0.96$ and $\alpha=1.69$.
To better capture the unrelated nature of the processors as well as data locality issues, we generate the task processing times in each processor in a processor-task correlated way, extending on the model of \cite{Hariri1991}. Specifically, the processing times $\{p_{i,k,j}\}_{i\in\mathcal{P}}$ of each map task are equal to $b_j a_{j,i}$ plus some noise selected u.a.r. from $[0,10]$, where $b_j$ and $a_{j,i}$ are selected u.a.r. from $[1,10]$, for each job $j\in\mathcal{J}$ and each processor $i\in \mathcal{P}$. The processing time of each reduce task, taking into account that is practically larger, is set to $3b_j a_{j, i}$ plus some noise selected u.a.r. from $[0,10]$. In this context, we simulate both Algorithm~\ref{alg:mrs} and Fast-MR by running 10 different trials for each possible number of jobs. Since in various applications a MapReduce computation is performed within a single round, we also simulate Algorithm~\ref{alg:mrs} in the single-round case, called \textsc{Single-Round MRS} and compare it against Fast-MR. Note that in the latter case, we fix $\alpha = 1.65, \delta= 0.80$ according to Corollary 1. The instances and the results are available at \url{http://www.corelab.ntua.gr/~opapadig/mrrounds/}.

\input plots.tex
	
In Figures~(i)-(ii), we note that Algorithm~\ref{alg:mrs} outperforms the Fast-MR heuristic, for any simulated number of jobs. More specifically, the empirical approximation ratio of Fast-MR, ranges from $3.32$ to $4.30$, while the ratio of Algorithm~\ref{alg:mrs} ranges from $2.57$ to $3.68$. More interestingly, the gap between the performance guarantee of the two algorithms is growing as the number of jobs is increasing: For $n=5$ jobs the average ratios of the algorithms Algorithm~\ref{alg:mrs} and Fast-MR are $3.43$ and $3.72$, while for $n=50$, the average ratio converges to $2.71$ and $3.62$, respectively. Over all trials, we can see that Algorithm~\ref{alg:mrs} produces up to $28.4\%$ better solutions.
In Figures~(iii)-(iv), we note that \textsc{Single-round MRS} also outperforms Fast-MR, producing up to $36.7 \%$ better solutions. Similarly to Algorithm~\ref{alg:mrs}, its empirical approximation ratio ranges from $2.25$ to $3.78$ (vs. the ratio of Fast-MR which ranges from $2.94$ to $4.44$), while the gap against the approximation ratio of Fast-MR increases as the number of jobs increasing (e.g., for $n=50$, \textsc{Single-round MRS} achieves ratio $2.37$, while Fast-MR $3.40$). Note that, the empirical approximation ratios in both multi-round and single-round cases of our algorithm is far from our theoretical worst-case approximation guarantees.




\newpage
\begin{thebibliography}{10}
%
\bibitem{ASSU13}
F.N. Afrati, A.~Das Sarma, S.~Salihoglu, and J.D. Ullman.
\newblock Upper and Lower Bounds on the Cost of a MapReduce Computation.
\newblock {\em VLDB}, 6(4):277--288, 2013.
%
%
\bibitem{AJRSU15}
F. Afrati, M. Joglekar, C. Ré, S. Salihoglu, and J.D. Ullman.
\newblock GYM: A multiround join algorithm in mapreduce.
\newblock {\em arXiv preprint arXiv:1410.4156}, 2014.
%
%
\bibitem{AspnesAFPW97}
J.~Aspnes, Y.~Azar, A.~Fiat, S.~Plotkin, and O.~Waarts.
 \newblock On-line Routing of Virtual Circuits with Applications to Load Balancing and Machine Scheduling.
 \newblock {\em Journal of the ACM}, 44(3):486--504, 1997.
%
%
\bibitem{ChenKL12}
F.~Chen, M.~S. Kodialam, and T.~V. Lakshman.
\newblock {\em Joint scheduling of processing and shuffle phases in mapreduce
  systems.}
\newblock In {IEEE} Proceedings of the 31st International Conference on
  Computer Communications(INFOCOM), pages 1143--1151, 2012.
%
\bibitem{CorreaSV12}
J.~R.~Correa, M.~Skutella, J.~Verschae.
\newblock The power of preemption on unrelated machines and applications to
  scheduling orders.
\newblock {\em Mathematics of Operations Research}, 37(2):379--398, 2012.
%
\bibitem{DEAN}
J.~Dean and S.~Ghemawat.
\newblock {\em Mapreduce: Simplified data processing on large clusters.}
\newblock In Proceedings of the 6th Symposium on Operating System Design
  and Implementation, pages 137--150, 2004.
%
\bibitem{EastmanEI64}
W.L. Eastman, S. Even, I.M. Iaacs.
\newblock Bounds for the optimal scheduling of $n$ jobs on $m$ processors.
\newblock {\em Management Science.} 11:268-279, 1964.
%
\bibitem{FM15}
D. Fotakis, I. Milis, O. Papadigenopoulos, E. Zampetakis and G. Zois.
\newblock {\em Scheduling MapReduce Jobs and Data Shuffle on Unrelated Processors.}
\newblock Proceedings of the 14th International Symposium on Experimental Algorithms (SEA), pages 137--150, 2015.
%
%
%
%
\bibitem{Graham69}
R.L. Graham.
\newblock Bounds on multiprocessing timing anomalies.
\newblock {\em SIAM journal on Applied Mathematics}, 17(2):416-429, 1969.
%
\bibitem{HallSSW97}
L.~A. Hall, A.S. Schulz, D.~B. Shmoys, and J.~Wein.
\newblock Scheduling to minimize average completion time: Off-line and on-line
  approximation algorithms.
\newblock {\em Mathematics of Operations Research}, 22:513--544, 1997.
%
\bibitem{Hariri1991}
A.~M.~Hariri, and C.~N. Potts.
\newblock Heuristics for scheduling unrelated parallel machines.
\newblock {\em Computers and Operations Research}, 18(3):323--331, 1991.
%
\bibitem{ImM15}
S. Im, and B. Moseley.
\newblock {\em Brief Announcement: Fast and Better Distributed MapReduce Algorithms for k-Center Clustering.}
\newblock Proceedings of the 27th ACM on Symposium on Parallelism in Algorithms and Architectures (SPAA), pages 65--67, 2015.
%
\bibitem{KarloffSV10}
H. Karloff, S. Suri, S. Vassilvitskii.
\newblock {\em A Model of Computation for MapReduce.}
\newblock Proceedings of the 21st Annual ACM-SIAM Symposium on Discrete Algorithms (SODA), pages 263--285, 2010.
%
\bibitem{KumarMPS09}
V.~S.~A. Kumar, M.~V. Marathe, S.~Parthasarathy, and A.~Srinivasan.
\newblock Scheduling on unrelated machines under tree-like precedence constraints.
\newblock {\em Algorithmica}, 55(1):205--226, 2009.
%
\bibitem{MoseleySPAA13}
R.~Kumar, B.~Moseley, S.~Vassilvitskii, and A.~Vattani.
\newblock {\em Fast greedy algorithms in mapreduce and streaming.}
\newblock In Proceedings of the 25th {ACM} Symposium on Parallel Algorithms and Architectures (SPAA), pages 1--10, 2013.
%
\bibitem{LinV92}
J. Lin, J. ~S. Vitter.
\newblock {\em $\epsilon$-Approximations with Minimum Packing Constraint Violation.}
\newblock Proceedings of the 24th annual ACM Symposium on Theory of Computing (STOC), pages 771--782, 1992.
%
%
\bibitem{MastrolilliS11}
M.~Mastrolilli and O.~Svensson.
\newblock Hardness of approximating flow and job shop scheduling problems.
\newblock {\em Journal of the {ACM}}, 58(5):20, 2011.
%
\bibitem{MoseleyDKS11}
B.~Moseley, A.~Dasgupta, R.~Kumar, and T.~Sarl{\'o}s.
\newblock {\em On scheduling in map-reduce and flow-shops.}
\newblock In Proceedings of the 23rd {ACM} Symposium on Parallel
  Algorithms and Architectures (SPAA), pages 289--298, 2011.
%
\bibitem{QueyranneS06}
M. Queyranne, and A.S. Schulz.
\newblock Approximation bounds for a general class of precedence constrained parallel machine scheduling problems.
\newblock {\em SIAM Journal on Computing}, 35(5):1241-1253, 2006.
%
\bibitem{QUE}
M. Queyranne.
\newblock Structure of a simple scheduling polyhedron.
\newblock \emph{Mathematical Programming}, 58(1): 263–285, 1993.
%
%
%
%
\bibitem{ST93}
D.B. Shmoys and {\'E}.~Tardos.
\newblock An approximation algorithm for the generalized assignment problem.
\newblock {\em Mathematical Programming}, 62:461--474, 1993.
%
\bibitem{ShmoysSW94}
D.~B. Shmoys, C.~Stein, and J.~Wein.
\newblock Improved approximation algorithms for shop scheduling problems.
\newblock {\em {SIAM} Journal on Computing}, 23(3):617--632, 1994.
%
%
%
\bibitem{YooS11}
D.-J. Yoo and K.~M. Sim.
\newblock {\em A comparative review of job scheduling for mapreduce.}
\newblock In {IEEE} Proceedings of the International Symposium on Cloud
  Computing and Intelligece Systems, pages 353--358, 2011.
\end{thebibliography}
\end{document}